\documentclass[aps,prd,twocolumns,eqsecnum,preprintnumbers,showpacs,amsmath,amssymb]
{revtex4}

\usepackage{graphicx}% Include figure files
\usepackage{dcolumn}% Align table columns on decimal point
\usepackage{bm}% bold math

\begin{document}

\title{HOW TO DEMONSTRATE A POSSIBLE EXISTENCE \\ OF A MASS GAP IN QCD}

\author{V. Gogokhia}
\email[]{gogohia@rmki.kfki.hu}

\affiliation{HAS, CRIP, RMKI, Depart. Theor. Phys., Budapest 114,
P.O.B. 49, H-1525, Hungary}

\date{\today}
\begin{abstract}
We propose to realize a mass gap in QCD by not imposing the
transversality condition on the full gluon self-energy, while
preserving the color gauge invariance condition for the full gluon
propagator. This is justified by the nonlinear and nonperturbative
dynamics of QCD. None of physical observables/processes in
low-energy QCD will be directly affected by such a temporary
violation of color gauge invariance/symmetry. No
truncations/approximations and no special gauge choice are made
for the regularized skeleton loop integrals, contributing to the
full gluon self-energy, which enters the Schwinger-Dyson equation
for the full gluon propagator. In order to make the existence of a
mass gap perfectly clear the corresponding subtraction procedure
is introduced. All this allows one to establish the general
structure of the full gluon propagator and the corresponding gluon
Schwinger-Dyson equation in the presence of a mass gap. It is
mainly generated by the nonlinear interaction of massless gluon
modes. The physical meaning of the mass gap is to be responsible
for the large-scale (low-energy/momentum), i.e., nonperturbative
structure of the true QCD vacuum. In the presence of a mass gap
two different types of solutions for the full gluon propagator are
possible. The massive solution leads to an effective gluon mass,
which depends on the gauge-fixing parameter explicitly. This
solution becomes smooth at small gluon momentum in the Landau
gauge. The general iteration solution is always severely singular
at small gluon momentum, i.e., the gluons remain massless, and
this does not depend on the gauge choice.

\end{abstract}



\maketitle

\section{Introduction}

Today there is no doubt left that color confinement and other
dynamical effects, such as spontaneous breakdown of chiral
symmetry, bound-state problems, etc., being essentially
nonperturbative (NP) effects, are closely related to the
large-scale (low-energy/momentum) structure of the true QCD ground
state and vice-versa \cite{1,2} (and references therein). The
perturbation theory (PT) methods in general fail to investigate
them. If QCD itself is a confining theory then a characteristic
scale has to exist. It should be directly responsible for the
above-mentioned structure of the true QCD vacuum in the same way
as $\Lambda_{QCD}$ is responsible for the nontrivial perturbative
dynamics there (scale violation, asymptotic freedom (AF)
\cite{3}).

The Lagrangian of QCD \cite{3,4} does not contain explicitly any
of the mass scale parameters which could have a physical meaning
even after the corresponding renormalization program is performed.
The main goal of this paper is to show how a characteristic scale
(the mass gap, for simplicity) responsible for the NP dynamics in
the infrared (IR) region may explicitly appear in QCD. This
becomes an imperative especially after Jaffe and Witten have
formulated their theorem "Yang-Mills Existence And Mass Gap"
\cite{5}. We will show that the mass gap is dynamically generated
mainly due to the nonlinear (NL) interaction of massless gluon
modes.

The propagation of gluons is one of the main dynamical effects in
the true QCD vacuum. It is described by the corresponding quantum
equation of motion, the so-called Schwinger-Dyson (SD) equation
\cite{1} (and references therein) for the full gluon propagator.
The importance of this equation is due to the fact that its
solutions reflect the quantum-dynamical structure of the true QCD
ground state. The color gauge structure of this equation is the
main subject of our investigation in order to find a way how to
realize a mass gap in QCD. Also we will discuss at least two
possible types of solutions of the gluon SD equation in the
presence of a mass gap, making no approximations/truncations and
no special gauge choice for the skeleton loop integrals
contributing to it. So they can be considered as the
generalizations of the explicit solutions because the latter ones
are necessarily based on the above-mentioned specific
approximations/truncations schemes.

\section{QED}

It is instructive to begin with a brief explanation why a mass gap
does not occur in quantum electrodynamics (QED). The photon SD
equation can be symbolically written down as follows:

\begin{equation}
D(q) = D^0(q) + D^0(q) \Pi(q) D(q),
\end{equation}
where we omit, for convenience, the dependence on the Dirac
indices, and $D^0(q)$ is the free photon propagator. $\Pi(q)$
describes the electron skeleton loop contribution to the photon
self-energy (the so-called vacuum polarization tensor).
Analytically it looks

\begin{equation}
\Pi(q) \equiv \Pi_{\mu\nu}(q) = - g^2 \int {i d^4 p \over (2
\pi)^4} Tr [\gamma_{\mu} S(p-q) \Gamma_{\nu}(p-q, q)S(p)],
\end{equation}
where $S(p)$ and $\Gamma_{\mu}(p-q,q)$ represent the full electron
propagator and the full electron-photon vertex, respectively. Here
and everywhere below the signature is Euclidean, since it implies
$q_i \rightarrow 0$ when $q^2 \rightarrow 0$ and vice-versa. This
tensor has the dimensions of a mass squared, and therefore it is
quadratically divergent. To make the formal existence of a mass
gap (the quadratically divergent constant, so having the dimension
of a mass squared) perfectly clear, let us now, for simplicity,
subtract its value at zero. One obtains

\begin{equation}
\Pi^s(q) \equiv \Pi^s_{\mu\nu}(q) = \Pi_{\mu\nu}(q) -
\Pi_{\mu\nu}(0) = \Pi_{\mu\nu}(q) - \delta _{\mu\nu}\Delta^2
(\lambda).
\end{equation}
The explicit dependence on the dimensionless ultraviolet (UV)
regulating parameter $\lambda$ has been introduced into the mass
gap $\Delta^2(\lambda)$, given by the integral (2.2) at $q^2=0$,
in order to assign a mathematical meaning to it. In this
connection, a few remarks are in order in advance. The dependence
on $\lambda$ (when it is not shown explicitly) is assumed in all
divergent integrals here and below in the case of the gluon
self-energy as well (see next section). This means that all the
expressions are regularized (including photon/gluon propagator),
and we can operate with them as with finite quantities. $\lambda$
should be removed on the final stage only after performing the
corresponding renormalization program (which is beyond the scope
of the present investigation, of course). Whether the regulating
parameter $\lambda$ has been introduced in a gauge-invariant way
(though this always can be achieved) or not, and how it should be
removed is not important for the problem if a mass gap can be
"released/liberated" from the corresponding vacuum. We will show
in the most general way (not using the PT and no special gauge
choice will be made) that this impossible in QED and might be
possible in QCD.

The tensor structure of the subtracted photon self-energy can be
written as follows:

\begin{equation}
\Pi^s_{\mu\nu}(q) = T_{\mu\nu}(q) q^2 \Pi^s_1(q^2) + q_{\mu}
q_{\nu}(q) \Pi^s_2(q^2),
\end{equation}
where both invariant functions $\Pi^s_n(q^2)$ at $n=1,2$ are, by
definition, dimensionless and regular at small $q^2$, since
$\Pi^s(0) =0$; otherwise they remain arbitrary. From this relation
it follows that $\Pi^s(q) = O(q^2)$, i.e., it is always of the
order $q^2$. Also, here and everywhere below

\begin{equation}
T_{\mu\nu}(q)=\delta_{\mu\nu}-q_{\mu} q_{\nu} / q^2 =
\delta_{\mu\nu } - L_{\mu\nu}(q).
\end{equation}
Taking into account the subtraction (2.3), the photon SD equation
becomes

\begin{equation}
D(q) = D^0(q) + D^0(q) \Pi^s(q) D(q) + D^0(q) \Delta^2(\lambda)
D(q).
\end{equation}
Its subtracted part can be summed up into the geometric series, so
one has

\begin{equation}
D(q) = \tilde{D}^0(q) + \tilde{D}^0(q) \Delta^2(\lambda) D(q),
\end{equation}
where the modified photon propagator is

\begin{equation}
\tilde{D}^0(q) = {D^0(q) \over  1 - \Pi^s(q) D^0(q)}= D^0(q) +
D^0(q) \Pi^s(q) D^0(q) - D^0(q)\Pi^s(q)D^0(q) \Pi^s(q)D^0(q) + ...
\ .
\end{equation}
Since $\Pi^s(q) = O(q^2)$ and $D^0(q) \sim (q^2)^{-1}$, the IR
singularity of the modified photon propagator is determined by the
IR singularity of the free photon propagator, i.e.,
$\tilde{D}^0(q) = O (D^0(q))$ with respect to the behavior at
small photon momentum (in Eqs. (2.6), (2.7) and (2.8) we again
omit the tensor indices, for simplicity).

Similar to the subtracted photon self-energy, the photon
self-energy (2.2) in terms of independent tensor structures is

\begin{equation}
\Pi_{\mu\nu}(q) = T_{\mu\nu} q^2 \Pi_1(q^2) + q_{\mu} q_{\nu}
\Pi_2(q^2),
\end{equation}
where both invariant functions $\Pi_n(q^2)$ at $n=1,2$ are
dimensionless and remain arbitrary. Due to the transversality of
the photon self-energy

\begin{equation}
q_{\mu} \Pi_{\mu\nu}(q) =q_{\nu} \Pi_{\mu\nu}(q) =0,
\end{equation}
which comes from the current conservation condition in QED, one
then has $\Pi_2(q^2) =0$, i.e., this tensor should be purely
transversal. On the other hand, from the subtraction (2.3) and
transversality condition (2.10) it follows that

\begin{equation}
\Pi_2^s(q^2) = -(\Delta^2(\lambda)/ q^2).
\end{equation}
However, this is impossible, since $\Pi_2^s(q^2)$ is a regular
function of $q^2$, so the mass gap should be zero and consequently
$\Pi_2^s(q^2)=0$ as well, i.e.,

\begin{equation}
\Pi_2^s(q^2) = \Delta^2(\lambda) =0.
\end{equation}
This means that the photon self-energy coincides with its
subtracted counterpart and both of them are purely transversal.
Moreover, this means that the photon self-energy does not have a
pole in its invariant function $\Pi_1(q^2)= \Pi^s_1(q^2)$. As
mentioned above, in obtaining these results neither the PT has
been used nor a special gauge has been chosen. So there is no
place for quadratically divergent constants in QED, while
logarithmic divergence still can be present in the invariant
function $\Pi_1(q^2)= \Pi_1^s(q^2)$. It is to be included into the
electric charge through the corresponding renormalization program
(for these detailed gauge-invariant derivations explicitly done in
lower order of the PT see Refs. \cite{4,6,7,8,9}).

In fact, the current conservation condition (2.10) lowers the
quadratical divergence of the corresponding integral (2.2) to a
logarithmic one. That is the reason why in QED logarithmic
divergences survive only. Thus in QED there is no mass gap and the
relevant photon SD equation is shown in Eq. (2.8), simply
identifying the full photon propagator with its modified
counterpart. In QED we should replace $\Pi(q)$ by its subtracted
counterpart $\Pi^s(q)$ from the very beginning ($\Pi(q)
\rightarrow \Pi^s(q)$), totally discarding the quadratically
divergent constant $\Delta^2(\lambda)$ from all the equations and
relations. The current conservation condition for the photon
self-energy (2.10), i.e., its transversality, and for the full
photon propagator $q_{\mu}q_{\nu}D_{\mu\nu}(q) = i\xi$, where
$\xi$ is the gauge-fixing parameter, are consequences of gauge
invariance. They should be maintained at every stage of the
calculations, since the photon is a physical state. In other
words, at all stages the current conservation plays a crucial role
in extracting physical information from the $S$-matrix elements in
QED. For example, if some QED process includes the full photon
propagator, then the corresponding $S$-matrix element is
proportional to the combination $j^{\mu}_1 (q)D_{\mu\nu}(q)
j^{\nu}_2(q)$. The current conservation condition $j^{\mu}_1 (q)
q_{\mu} = j^{\nu}_2(q)q_{\nu} =0$ implies that the unphysical
(longitudinal) component of the full photon propagator does not
change the physics of QED, i.e., only its physical (transversal)
component is important. In its turn this means that the
transversality condition imposed on the photon self-energy is
important, since $\Pi_{\mu\nu}(q)$ itself is a correction to the
amplitude of the physical process, for example such as
electron-electron scattering.

\section{QCD}

Due do color confinement in QCD the gluon is not a physical state.
Still, color gauge invariance should also be preserved, so the
color current conservation takes place in QCD as well. However, in
this theory it plays no role in the extraction of physical
information from the $S$-matrix elements for the corresponding
physical processes and quantities. So in QCD there is no such
physical amplitude to which the gluon self-energy may directly
contribute (for example, quark-quark/antiquark scattering is not a
physical process). The lesson which comes from QED is that if one
preserves the transversality of the photon self-energy at every
stage, then there is no mass gap. Thus, in order to realize a mass
gap in QCD, our proposal is not to impose the transversality
condition on the gluon self-energy, but preserving the color gauge
invariance condition for the full gluon propagator (see below). As
mentioned above, no QCD physics will be directly affected by this.
So color gauge symmetry will be violated at the initial stage (at
the level of the gluon self-energy) and will be restored at the
final stage (at the level of the full gluon propagator).

\subsection{Gluon SD equation}

The gluon SD equation symbolically is

\begin{equation}
D_{\mu\nu}(q) = D^0_{\mu\nu}(q) + D^0_{\mu\rho}(q) i
\Pi_{\rho\sigma}(q; D) D_{\sigma\nu}(q),
\end{equation}
where $D^0_{\mu\nu}(q)$ is the free gluon propagator.
$\Pi_{\rho\sigma}(q; D)$ is the gluon self-energy, and in general
it depends on the full gluon propagator due to the non-Abelian
character of QCD (see below). Thus the gluon SD equation is highly
NL, while the photon SD equation (2.1) is a linear one. In what
follows we omit the color group indices, since for the gluon
propagator (and hence for its self-energy) they are reduced to the
trivial $\delta$-function, for example $D^{ab}_{\mu\nu}(q) =
D_{\mu\nu}(q)\delta^{ab}$. Also, for convenience, we introduce $i$
into the gluon SD equation (3.1).

The gluon self-energy $\Pi_{\rho\sigma}(q; D)$ is the sum of a few
terms, namely

\begin{equation}
 \Pi_{\rho\sigma}(q; D)= - \Pi^q_{\rho\sigma}(q) -
\Pi^{gh}_{\rho\sigma}(q) + \Pi_{\rho\sigma}^t(D) +
\Pi_{(1)\rho\sigma}(q; D) + \Pi_{(2)\rho\sigma}(q; D) +
\Pi'_{(2)\rho\sigma}(q; D),
\end{equation}
where $\Pi^q_{\rho\sigma}(q)$ describes the skeleton loop
contribution due to quark degrees of freedom (it is an analog of
the vacuum polarization tensor in QED, see Eq. (2.2)), while
$\Pi^{gh}_{\rho\sigma}(q)$ describes the skeleton loop
contribution due to ghost degrees of freedom. Both skeleton loop
integrals do not depend on the full gluon propagator $D$, so they
represent the linear contribution to the gluon self-energy.
$\Pi_{\rho\sigma}^t(D)$ represents the so-called constant skeleton
tadpole term. $\Pi_{(1)\rho\sigma}(q; D)$ represents the skeleton
loop contribution, which contains the triple gluon vertices only.
$\Pi_{(2)\rho\sigma}(q; D)$ and $\Pi'_{(2)\rho\sigma}(q; D)$
describe topologically independent skeleton two-loop
contributions, which combine the triple and quartic gluon
vertices. The last four terms explicitly contain the full gluon
propagators in different powers, that is why they form the NL part
of the gluon self-energy. The explicit expressions for the
corresponding skeleton loop integrals \cite{10} (in which the
corresponding symmetry coefficients can be included) are of no
importance here. Let us note that like in QED these skeleton loop
integrals are in general quadratically divergent, and therefore
they should be regularized (see remarks above and below).

\subsection {A temporary violation of color gauge invariance/symmetry
(TVCGI/S)}

 The color gauge invariance condition for the gluon
self-energy (3.2) can be reduced to the three independent
transversality conditions imposed on it. It is well known that the
quark contribution can be made transversal independently of the
pure gluon contributions within any regularization scheme which
preserves gauge invariance, for example such as the dimensional
regularization method (DRM) \cite{3,4,8,9,11}. So, one has

\begin{equation}
q_{\rho} \Pi^q_{\rho\sigma}(q) =q_{\sigma} \Pi^q_{\rho\sigma}(q) =
0,
\end{equation}
indeed. In the same way the sum of the gluon contributions can be
done transversal by taking into account the ghost contribution, so
again one has

\begin{equation}
q_{\rho} \Bigl[ \Pi_{(1)\rho\sigma}(q; D) + \Pi_{(2)\rho\sigma}(q;
D) + \Pi'_{(2)\rho\sigma}(q; D) - \Pi^{gh}_{\rho\sigma}(q) \Bigr]
= 0.
\end{equation}
The role of ghost degrees of freedom is to cancel the unphysical
(longitudinal) component of gauge bosons in every order of the PT,
i.e., going beyond the PT and thus being general. The previous
relation just demonstrate this, since it contains the
corresponding skeleton loop integrals.

However, there is no such regularization scheme (preserving or not
gauge invariance) in which the transversality condition for the
constant skeleton tadpole term could be satisfied, i.e., $q_{\rho}
\Pi^t_{\rho\sigma}(D) = q_{\rho} \delta_{\rho\sigma} \Delta^2_t(D)
= q_{\sigma} \Delta^2_t(D) \neq 0$, indeed. This means that in any
NP approach the transversality condition imposed on the gluon
self-energy may not be valid, i.e., in general

\begin{equation}
q_{\rho} \Pi_{\rho\sigma}(q; D) =q_{\sigma} \Pi_{\rho\sigma}(q; D)
\neq 0.
\end{equation}
In the PT, when the full gluon propagator is always approximated
by the free one, the constant tadpole term is set to be zero
within the DRM \cite{8,11}, i.e., $\Pi^t_{\rho\sigma}(D^0) =0$. So
in the PT the transversality condition for the gluon self-energy
is always satisfied.

The relation (3.5) justifies our proposal not to impose the
transversality condition on the gluon self-energy. The special
role of the constant skeleton tadpole term in the NP QCD dynamics
should be emphasized. It explicitly violates the transversality
condition for the gluon self-energy (3.5). The second important
observation is that now ghosts themselves cannot automatically
provide the transversality of the gluon propagator in NP QCD.
However, this does not mean that we need no ghosts at all. Of
course, we need them in other sectors of QCD, for example in the
quark-gluon Ward-Takahashi identity, which contains the so-called
ghost-quark scattering kernel explicitly \cite{3}.

\subsection{Subtractions}

As we already know from QED, the regularization of the gluon
self-energy can be started from the subtraction its value at the
zero point (see, however, remarks below). Thus, quite similarly to
the subtraction (2.3), one obtains

\begin{equation}
\Pi^s_{\rho\sigma}(q; D) = \Pi_{\rho\sigma}(q; D) -
\Pi_{\rho\sigma}(0; D) = \Pi_{\rho\sigma}(q; D) -
\delta_{\rho\sigma} \Delta^2 (\lambda; D).
\end{equation}
Let us remind once more that for our purpose, namely to
demonstrate a possible existence of a mass gap $\Delta^2 (\lambda;
D)$ in QCD, it is not important how $\lambda$ has been introduced
and  how it should be removed at the final stage. The mass gap
itself is mainly generated by the nonlinear interaction of
massless gluon modes, slightly corrected by the linear
contributions coming from the quark and ghost degrees of freedom,
namely

\begin{equation}
\Delta^2 (\lambda; D)= \Pi^t(D) + \sum_a \Pi^a(0; D) =
\Delta^2_t(D) + \sum_a \Delta^2_a(0; D) ,
\end{equation}
where index "a" runs as follows: $a= -q, -gh, 1, 2, 2'$, and $-q,
\ - gh$ mean that both terms enter the above-mentioned sum with
minus sign (here, obviously, the tensor indices are omitted). In
these relation all the divergent constants $\Pi^t(D)$ and
$\Pi^a(0; D)$, having the dimensions of a mass squared, are given
by the corresponding skeleton loop integrals at $q^2=0$. Thus
these constants summed up into the mass gap squared (3.8) cannot
be discarded like in QED, since the transversality condition for
the gluon self-energy is not satisfied, see Eq. (3.5). In other
words, in QCD in general the quadratical divergences of the
corresponding loop integrals cannot be lowered to logarithmic
ones, and therefore the mass gap (3.7) should be explicitly taken
into account in this theory. The transversality condition for the
gluon self-energy can be satisfied partially, i.e., if one imposes
it on quark and gluon (along with ghost) degrees of freedom as it
follows from above. Then the mass gap is to be reduced to
$\Pi^t(D)$, since all other constants $\Pi^a(0;D)$ can be
discarded in this case (see Eq. (3.7)). However, we will stick to
our proposal not to impose the transversality condition on the
gluon self-energy, and thus to deal with the mass gap on account
of all possible contributions.

The subtracted gluon self-energy

\begin{equation}
\Pi^s_{\rho\sigma}(q; D) \equiv \Pi^s(q; D) = \sum_a \Pi^s_a(q; D)
\end{equation}
is free from the tadpole contribution, because $\Pi^s_t(D) =
\Pi_t(D)- \Pi_t(D)=0$, by definition, at any $D$, while in the
gluon self-energy it is explicitly present through the mass gap
(see Eqs. (3.7) and (3.6)). The general decomposition of the
subtracted gluon self-energy into the independent tensor
structures can be written down as follows:

\begin{equation}
\Pi^s_{\rho\sigma}(q; D) = T_{\rho\sigma}(q) q^2 \Pi(q^2; D) +
q_{\rho} q_{\sigma} \tilde{\Pi}(q^2; D),
\end{equation}
where both invariant functions $\Pi(q^2; D)$ and $\tilde{\Pi}(q^2;
D)$ are dimensionless and regular at small $q^2$. Since the
subtracted gluon self-energy does not contain the tadpole
contribution, we can now impose the color current conservation
condition on it, i.e., to put

\begin{equation}
q_{\rho} \Pi^s_{\rho\sigma}(q; D)= q_{\sigma}
\Pi^s_{\rho\sigma}(q; D) = 0,
\end{equation}
which implies $\tilde{\Pi}(q^2; D) = 0$, so that the subtracted
gluon self-energy finally becomes purely transversal

\begin{equation}
\Pi^s_{\rho\sigma}(q; D) = T_{\rho\sigma}(q) q^2 \Pi(q^2; D),
\end{equation}
and it is always of the order $q^2$ at any $D$, since the
invariant function $\Pi(q^2; D)$ is regular at small $q^2$ at any
$D$. Thus the subtracted quantities are free from the quadratic
divergences, but logarithmic ones can be still present in
$\Pi(q^2; D)$ like in QED.

\subsection{General structure of the gluon SD equation}

Our strategy is not to impose the transversality condition on the
gluon self-energy in order to realize a mass gap despite whether
or not the tadpole term is explicitly present. To show that this
works, it is instructive to substitute the subtracted gluon
self-energy (3.9) (and not its transversal part (3.11)) into the
initial gluon SD equation (3.1), on account of the subtraction
(3.6). Then one obtains

\begin{equation}
D_{\mu\nu}(q) = D^0_{\mu\nu}(q) + D^0_{\mu\rho}(q)i[
T_{\rho\sigma}(q) q^2 \Pi(q^2; D) + q_{\rho}q_{\sigma}
\tilde{\Pi}(q^2; D)]D_{\sigma\nu}(q) + D^0_{\mu\sigma}(q)i
\Delta^2(\lambda; D) D_{\sigma\nu}(q).
\end{equation}
Let us now introduce the general tensor decompositions of the full
and auxiliary free gluon propagators
$D_{\mu\nu}(q)=i[T_{\mu\nu}(q) d(q^2) +
L_{\mu\nu}(q)d_1(q^2)](1/q^2)$ and

\begin{equation}
D^0_{\mu\nu}(q)=i[ T_{\mu\nu}(q) + L_{\mu\nu}(q) d_0(q^2)](1/q^2),
\end{equation}
respectively. The form factor $d_0(q^2)$ introduced into the
unphysical part of the auxiliary free gluon propagator
$D^0_{\mu\nu}(q)$ is needed in order to explicitly show that the
longitudinal part of the subtracted gluon self-energy
$\tilde{\Pi}(q^2; D)$ plays no role. The color gauge invariance
condition imposed on the full gluon propagator

\begin{equation}
q_{\mu}q_{\nu}D_{\mu\nu}(q) = i \xi,
\end{equation}
implies $d_1(q^2) = \xi$, so that the full gluon propagator
becomes

\begin{equation}
D_{\mu\nu}(q) = i \left\{ T_{\mu\nu}(q) d(q^2) + \xi L_{\mu\nu}(q)
\right\} {1 \over q^2}.
\end{equation}
Substituting all these decompositions into the gluon SD equation
(3.12), one obtains

\begin{equation}
d(q^2) = {1 \over 1 + \Pi(q^2; D) + (\Delta^2(\lambda; D) / q^2)},
\end{equation}
and

\begin{equation}
d_0(q^2) = {\xi \over 1 - \xi [\tilde{\Pi}(q^2; D) +
(\Delta^2(\lambda; D) / q^2)]}.
\end{equation}
However, the auxiliary free gluon propagator defined in Eqs.
(3.13) and (3.17) is to be equivalently replaced as follows:

\begin{equation}
D^0_{\mu\nu}(q) \Longrightarrow D^0_{\mu\nu}(q) + i \xi
L_{\mu\nu}(q) d_0(q^2) \Bigl[ \tilde{\Pi}(q^2; D) + {
\Delta^2(\lambda; D) \over q^2} \Bigr] {1 \over q^2},
\end{equation}
where $D^0_{\mu\nu}(q)$ in the right-hand-side is the standard
free gluon propagator, i.e.,

\begin{equation}
D^0_{\mu\nu}(q) = i \left\{ T_{\mu\nu}(q) + \xi L_{\mu\nu}(q)
\right\} {1 \over q^2}.
\end{equation}
Then the gluon SD equation in the presence of the mass gap (3.12),
on account of the explicit expression for the auxiliary free gluon
form factor (3.17), and doing some tedious algebra, is also to be
equivalently replaced as follows:

\begin{eqnarray}
D_{\mu\nu}(q) &=& D^0_{\mu\nu}(q) + D^0_{\mu\rho}(q)i
T_{\rho\sigma}(q) q^2 \Pi(q^2; D) D_{\sigma\nu}(q) \nonumber\\
&+& D^0_{\mu\sigma}(q)i \Delta^2(\lambda; D) D_{\sigma\nu}(q) + i
\xi^2 L_{\mu\nu}(q) { \Delta^2(\lambda; D) \over q^4}.
\end{eqnarray}
Here and below $D^0_{\mu\nu}(q)$ is the free gluon
propagator(3.19). The gluon SD equation (3.20) does not depend on
$d_0(q^2)$ and $\tilde{\Pi}(q^2; D)$, i.e., they played their role
and then retired from the scene. So, our derivation explicitly
shows that the longitudinal part of the subtracted gluon
self-energy $\tilde{\Pi}(q^2; D)$ plays no role and can be put to
zero without loosing generality, and thus making the subtracted
gluon self-energy purely transversal in accordance with Eq.
(3.11).

Using now the explicit expression for the free gluon propagator
(3.19) this equation can be further simplified to

\begin{equation}
D_{\mu\nu}(q) = D^0_{\mu\nu}(q) - T_{\mu\sigma}(q) \Bigl[\Pi(q^2;
D) + { \Delta^2(\lambda; D) \over q^2} \Bigr] D_{\sigma\nu}(q).
\end{equation}
It is easy to check that the full gluon propagator satisfies the
color gauge invariance condition (3.14), indeed. So the full gluon
propagator is the expression (3.15) with the full gluon form
factor given in Eq. (3.16), which obviously satisfies Eq. (3.21).
The only price we have paid by violating color gauge invariance is
the gluon self-energy, while the full and free gluon propagators
and the subtracted gluon self-energy always satisfy it. Let us
emphasize that the expression for the full gluon form factor shown
in the relation (3.16) cannot be considered as the formal solution
for the full gluon propagator, since both the mass gap
$\Delta^2(\lambda; D)$ and the invariant function $\Pi(q^2; D)$
depend on $D$ themselves. Here it is worth noting in advance that
from above it is almost clear that if one begins with the UV
renormalization program, then the information on the mass gap will
be totally lost. In this case instead of the regularized gluon
self-energy its subtracted regularized counterpart comes into the
play. In other words, in the PT limit $\Delta^2(\lambda; D)=0$ one
recovers the standard gluon SD equation, and the gluon self-energy
coincides with its subtracted counterpart like in QED.

Thus, we have established the general structure of the full gluon
propagator (see Eqs. (3.15) and (3.16)) and the corresponding
gluon SD equation (3.21) (which is equivalent to Eq. (3.12)) in
the presence of a mass gap.

\section{Massive solution}

An immediate consequence of the explicit presence of the mass gap
in the full gluon propagator is that a massive-type solution for
it becomes possible. In other words, in this case the gluon may
indeed acquire an effective mass. From Eq. (3.16) it follows that

\begin{equation}
{ 1 \over q^2} d(q^2) = {1 \over q^2 + q^2 \Pi(q^2; \xi) +
\Delta^2(\lambda, \xi)},
\end{equation}
where instead of the dependence on $D$ the dependence on $\xi$ is
explicitly shown. The full gluon propagator (3.15) may have a
pole-type solution at the finite point if and only if the
denominator in Eq. (4.1) has a zero at this point $q^2 = - m^2_g$
(Euclidean signature), i.e.,

\begin{equation}
- m^2_g  - m^2_g \Pi(-m^2_g; \xi) + \Delta^2(\lambda, \xi)=0,
\end{equation}
where $m^2_g \equiv m^2_g(\lambda, \xi)$ is an effective gluon
mass, and the previous equation is a transcendental equation for
its determination. Excluding the mass gap, one obtains that the
denominator in the full gluon propagator becomes $q^2 + q^2
\Pi(q^2; \xi) + \Delta^2(\lambda, \xi) = q^2 + m^2_g + q^2
\Pi(q^2; \xi) + m^2_g \Pi(-m^2_g; \xi)$. Let us now expand
$\Pi(q^2; \xi)$ in a Taylor series near $m^2_g$:

\begin{equation}
\Pi(q^2; \xi) = \Pi(-m^2_g; \xi) + (q^2 + m^2_g) \Pi'(-m^2_g; \xi)
+ O \Bigl( (q^2 + m^2_g)^2 \Bigr).
\end{equation}
Substituting this expansion into the previous relation and after
doing some tedious algebra, one obtains $q^2 + m^2_g + q^2
\Pi(q^2; \xi) + m^2_g \Pi(-m^2_g; \xi)= (q^2 + m^2_g)[1 +
\Pi(-m^2_g; \xi) - m^2_g \Pi'(-m^2_g; \xi)] [ 1 + \Pi^R(q^2;
\xi)]$, where $\Pi^R(q^2; \xi)= 0$ at $q^2=-m^2_g$ (otherwise it
remains arbitrary). Thus the full gluon propagator (3.15) now
looks

\begin{equation}
D_{\mu\nu}(q) = i T_{\mu\nu}(q) {Z_3 \over (q^2 + m^2_g) [ 1 +
\Pi^R(q^2; m^2_g)]} + i \xi L_{\mu\nu}(q) {1 \over q^2},
\end{equation}
where, for future purpose, in the invariant function $\Pi^R(q^2;
m^2_g)$ instead of $\xi$ we introduced the dependence on the gluon
effective mass squared $m_g^2$ which depends on $\xi$ itself. The
gluon renormalization constant is $Z_3 = [1 + \Pi(-m^2_g; \xi) -
m^2_g \Pi'(-m^2_g; \xi)]^{-1}$. In the formal PT limit
$\Delta^2(\lambda, \xi) =0$, an effective gluon mass is also zero,
$m_g^2(\lambda, \xi) =0$, as it follows from Eq. (4.2). So an
effective gluon mass is the NP effect. At the same time, it cannot
be interpreted as the "physical" gluon mass, since it remains
explicitly gauge-dependent quantity. The gluon renormalization
constant in this limit becomes a standard one, namely $[1 + \Pi(0;
\xi)]^{-1}$. The massive-type solution (4.4) becomes smooth in the
IR ($q^2 \rightarrow 0$) in the Landau gauge $\xi=0$ only (the
ghosts now cannot guarantee the cancellation of the longitudinal
part of the full gluon propagator as mentioned above). In this
connection let us point out that Landau gauge smooth (even
vanishing in the IR) gluon propagator at the expense of more
singular (than the free one) in the IR ghost propagator has been
obtained and discussed (see, for example Refs. \cite{12,13} and
references therein). As mentioned above, however, these results
are necessarily based on different approximations/truncations for
the skeleton loop integrals contributing to the gluon self-energy.

\section{Iteration solution}

 In order to perform a formal iteration of
the gluon SD equation (3.21), much more convenient to address to
its "solution" for the full gluon form factor (3.16),
nevertheless, and rewrite it as follows:

\begin{equation}
d(q^2) = 1 - \Bigl[ \Pi(q^2; d) + {\Delta^2(\lambda; d) \over q^2}
\Bigr] d(q^2) = 1 - P(q^2; d) d(q^2),
\end{equation}
i.e., in the form of the corresponding transcendental (i.e., not
algebraic) equation suitable for the formal nonlinear iteration
procedure. Here we replace the dependence on $D$ by the equivalent
dependence on $d$. For future purposes, it is convenient to
introduce short-hand notations as follows:

\begin{eqnarray}
\Delta^2(\lambda; d=d^{(0)} + d^{(1)} + d^{(2)} + ... + d^{(m)}+
...
) &=& \Delta^2_m = \Delta^2 c_m(\lambda, \alpha, \xi, g^2), \nonumber\\
\Pi(q^2; d=d^{(0)} + d^{(1)}+d^{(2)}+ ... + d^{(m)} + ...) &=&
\Pi_m(q^2) = [ P_m(q^2) - (\Delta^2_m / q^2) ].
\end{eqnarray}
In these relations $\Delta^2_m$ are the auxiliary mass squared
parameters, while $\Delta^2$ is the mass gap itself (see, however,
remarks in Conclusions). The dimensionless constants $c_m$ via the
corresponding subscripts depend on which iteration for the gluon
form factor $d$ is actually done. They may depend on the
dimensionless coupling constant squared $g^2$, as well as on the
gauge-fixing parameter $\xi$. We also introduce the explicit
dependence on the dimensionless finite (slightly different from
zero) subtraction point $\alpha$, since the initial subtraction at
the zero point may be dangerous \cite{3}. The dependence of
$\Delta^2$ on all these parameters is not shown explicitly, and if
necessary can be restored any time. Let us also remind that all
the invariant functions $\Pi_m(q^2)$ are regular at small $q^2$.
If it were possible to express the full gluon form factor $d(q^2)$
in terms of these quantities then it would be the formal solution
for the full gluon propagator. In fact, this is nothing but the
skeleton loops expansion, since the regularized skeleton loop
integrals, contributing to the gluon self-energy, have to be
iterated. This is the so-called general iteration solution. No
truncations/approximations and no special gauge choice have been
made. This formal expansion is not a PT series. The magnitude of
the coupling constant squared and the dependence of the
regularized skeleton loop integrals on it is completely arbitrary.

It is instructive to describe the general iteration procedure in
some details. Evidently, $d^{(0)}=1$ and doing the first iteration
in Eq. (5.1), one thus obtains

\begin{equation}
d(q^2) = 1 - P_0(q^2) + ... = 1 + d^{(1)}(q^2) + ...,
\end{equation}
where obviously $d^{(1)}(q^2) = - P_0(q^2)$. Doing the second
iteration, one obtains

\begin{equation}
d(q^2) = 1 - P_1(q^2) [ 1 + d^{(1)}(q^2) ] + ... = 1 +
d^{(1)}(q^2) + d^{(2)}(q^2) + ...,
\end{equation}
where $d^{(2)}(q^2) = - d^{(1)}(q^2) - P_1(q^2) [ 1 - P_0(q^2)]$.
Doing the third iteration, one further obtains

\begin{equation}
d(q^2) = 1 - P_2(q^2) [ 1 + d^{(1)}(q^2) + d^{(2)}(q^2)] + ... = 1
+ d^{(1)}(q^2) + d^{(2)}(q^2) + d^{(3)}(q^2) + ...,
\end{equation}
where $d^{(3)}(q^2) = - d^{(1)}(q^2) - d^{(2)}(q^2) - P_2(q^2) [ 1
- P_1(q^2)(1 - P_0(q^2))]$, and so on for the next iterations.
Thus up to the third iteration, one finally obtains

\begin{equation}
d(q^2) = \sum_{m=0}^{\infty} d^{(m)}(q^2) = 1 - [\Pi_2(q^2) +
{\Delta^2_2 \over q^2}] \Bigl[ 1 - [\Pi_1(q^2) + {\Delta^2_1 \over
q^2}] [1 - \Pi_0(q^2) - {\Delta^2_0 \over q^2}] \Bigr] + ... \ .
\end{equation}
We restrict ourselves to the third iterated term, since this
already allows to show explicitly some general features of such
kind of the nonlinear iteration procedure.

\subsection{ Splitting/shifting procedure}

Doing some tedious algebra, the previous expression can be
rewritten as follows:

\begin{eqnarray}
d(q^2) &=& [1 - \Pi_2(q^2) + \Pi_1(q^2) \Pi_2(q^2) - \Pi_0(q^2)
\Pi_1(q^2)\Pi_2(q^2) + ...] \nonumber\\
&+& {1 \over q^2} [\Pi_2(q^2)\Delta^2_1 + \Pi_1(q^2)\Delta^2_2 -
\Pi_0(q^2) \Pi_1(q^2)\Delta^2_2 - \Pi_0(q^2) \Pi_2(q^2)\Delta^2_1
- \Pi_1(q^2) \Pi_2(q^2)\Delta^2_2 + ...] \nonumber\\
&-& {1 \over q^4} [\Pi_0(q^2) \Delta^2_1 \Delta^2_2 + \Pi_1(q^2)
\Delta^2_0 \Delta^2_2 + \Pi_2(q^2) \Delta^2_0 \Delta^2_1 + ...]
\nonumber\\
&-& {1 \over q^2} [\Delta^2_2 -  {\Delta^2_1 \Delta^2_2 \over q^2}
+ { \Delta^2_0 \Delta^2_1 \Delta^2_2 \over q^4} + ...],
\end{eqnarray}
so that  this formal expansion contains three different types of
terms. The first type are the terms which contain only different
combinations of $\Pi_m(q^2)$ (they are not multiplied by inverse
powers of $q^2$); the third type of terms contains only different
combinations of $(\Delta^2_m / q^2)$. The second type of terms
contains the so-called mixed terms, containing the first and third
types of terms in different combinations. The two last types of
terms are multiplied by the corresponding powers of $1/q^2$.
Evidently, such structure of terms will be present in each
iteration term for the full gluon form factor. However, any of the
mixed terms can be split exactly into the first and third types of
terms by keeping the necessary number of terms in the Taylor
expansions in powers of $q^2$ for $\Pi_m(q^2)$, which are regular
functions at small $q^2$. Thus the IR structure of the full gluon
form factor (which just is our primary goal to establish) is
determined not only by the third type of terms. It gains
contributions from the mixed terms as well.

Let us present the above-mentioned Taylor expansions as follows:

\begin{equation}
\Pi_m(q^2) = \Pi_m(0) + (q^2 / \mu^2) \Pi^{(1)}_m (0) + (q^2 /
\mu^2)^2 \Pi^{(2)}_m (0) + O_m(q^6),
\end{equation}
since for the third iteration we need to use the Taylor expansions
up to this order (here $\mu^2$ is some fixed mass squared (not to
be mixed up with the tensor index)). For example, the mixed term
$(1/ q^2) \Pi_2(q^2)\Delta^2_1$ should be split as

\begin{equation}
{\Delta^2_1 \over q^2}\Pi_2(q^2) = {\Delta^2_1 \over q^2} \Bigl[
\Pi_2(0) + (q^2 / \mu^2) \Pi^{(1)}_2 (0) + O(q^4) \Bigr] =
{\Delta^2_1 \over q^2} \Pi_2(0) + a_1\Pi^{(1)}_2 (0) +O(q^2).
\end{equation}
Here and everywhere below $a_m = (\Delta^2_m / \mu^2), \
m=0,1,2,3,...$ are the dimensionless constants. The first term now
is to be shifted to the third type of terms and combined with the
term $(-1/q^2)\Delta^2_2$, while the second term $a_1\Pi^{(1)}_2
(0) +O(q^2)$ is to be shifted to the first type of terms. All
other mixed terms of similar structure should be treated
absolutely in the same way. For the mixed term $(-1 / q^4)
\Pi_0(q^2) \Delta^2_1 \Delta^2_2$, one has

\begin{eqnarray}
- {\Delta^2_1 \Delta^2_2 \over q^4}\Pi_0(q^2) &=& -{\Delta^2_1
\Delta^2_2 \over q^4} \Bigl[ \Pi_0(0) + (q^2 / \mu^2) \Pi^{(1)}_0
(0) +  (q^2 / \mu^2)^2 \Pi^{(2)}_0 (0) + O(q^6) \Bigr]
\nonumber\\
&=& - {\Delta^2_1 \Delta^2_2 \over q^4} \Pi_0(0) - {\Delta^2_1
\over q^2} a_2 \Pi^{(1)}_0 (0) - a_1a_2\Pi^{(2)}_0 (0)- O(q^2).
\end{eqnarray}
Again the first and second terms should be shifted to the third
type of terms and combined with terms containing there the same
powers of $1/q^2$, while the last two terms should be shifted to
the first type of terms.

Similar to the Taylor expansion (5.8), one has

\begin{equation}
\Pi_m(q^2)\Pi_n (q^2)= \Pi_{mn} (q^2) = \Pi_{mn}(0) + (q^2 /
\mu^2) \Pi^{(1)}_{mn} (0) + (q^2 / \mu^2)^2 \Pi^{(2)}_{mn} (0) +
O_{mn}(q^6).
\end{equation}
Then, for example the mixed term $(-1/q^2) \Pi_0(q^2)
\Pi_1(q^2)\Delta^2_2$ can be split as

\begin{eqnarray}
-{ \Delta^2_2 \over q^2} \Pi_0(q^2) \Pi_1(q^2) &=& -{ \Delta^2_2
\over q^2} \Bigl[ \Pi_{01}(0) + (q^2 / \mu^2) \Pi^{(1)}_{01} (0) +
O(q^4) \Bigr] \nonumber\\
&=& -{ \Delta^2_2 \over q^2} \Pi_{01}(0) - a_2 \Pi^{(1)}_{01} (0)
+ O(q^2),
\end{eqnarray}
so again the first term should be shifted to the third type of
terms and combined with the terms containing the corresponding
powers of $1/q^2$, while other terms are to be shifted to the
first type of terms.

Completing this exact splitting/shifting procedure in the
expansion (5.7), one can in general represent it as follows:

\begin{equation}
d(q^2) =(\Delta^2 / q^2) B_1(\lambda, \alpha, \xi, g^2) + (
\Delta^2 / q^2)^2 B_2(\lambda, \alpha, \xi, g^2) + (\Delta^2 /
q^2)^3 B_3(\lambda, \alpha, \xi, g^2) + f_3(q^2) + ...,
\end{equation}
where we used notations (5.2), since the coefficients of the
above-used Taylor expansions depend in general on the same set of
parameters: $\lambda, \alpha, \xi, g^2$. The invariant function
$f_3(q^2)$ is dimensionless and regular at small $q^2$; otherwise
it remains arbitrary. The generalization on the next iterations is
almost obvious. Let us only note that in this case more terms in
the corresponding Taylor expansions should be kept "alive".

\subsection{The exact structure of the general iteration solution}

Substituting the generalization of the expansion (5.13) on all
iterations and omitting the tedious algebra, the general iteration
solution of the gluon SD equation (3.21) for the regularized full
gluon propagator (3.15) can be exactly decomposed as the sum of
the two principally different terms as follows:

\begin{eqnarray}
D_{\mu\nu}(q; \Delta^2) = D^{INP}_{\mu\nu}(q; \Delta^2)+
D^{PT}_{\mu\nu}(q) &=& i T_{\mu\nu}(q) {\Delta^2 \over (q^2)^2}
\sum_{k=0}^{\infty} (\Delta^2 / q^2)^k \sum_{m=0}^{\infty}
\Phi_{k,m}(\lambda, \alpha,
\xi, g^2) \nonumber\\
&+& i \Bigr[ T_{\mu\nu}(q) \sum_{m=0}^{\infty} A_m(q^2) + \xi
L_{\mu\nu}(q) \Bigl] {1 \over q^2},
\end{eqnarray}
where the superscript "INP" stands for the intrinsically NP part
of the full gluon propagator. We distinguish between the two terms
in Eq. (5.14) by the character of the corresponding IR
singularities and the explicit presence of the mass gap (see
below). Let us emphasize that the general problem of convergence
of the formally regularized series (5.14) is irrelevant here.
Anyway, the problem how to remove all types of the UV divergences
(overlapping \cite{14} and overall \cite{3,4,6,7,8,9}) is a
standard one. Our problem will be how to deal with severe IR
singularities due to their novelty and genuine NP character.
Fortunately, there already exists a well-elaborated mathematical
formalism for this purpose, namely the distribution theory (DT)
\cite{15} to which the DRM \cite{11} should be correctly
implemented (see also Ref. \cite{10}).

The INP part of the full gluon propagator is characterized by the
presence of severe power-type (or equivalently NP) IR
singularities $(q^2)^{-2-k}, \ k=0,1,2,3,...$. So these IR
singularities are defined as more singular than the power-type IR
singularity of the free gluon propagator $(q^2)^{-1}$, which thus
can be defined as the PT IR singularity. The INP part depends only
on the transversal degrees of freedom of gauge bosons. Though its
coefficients $\Phi_{k,m}(\lambda, \alpha, \xi, g^2)$ may
explicitly depend on the gauge-fixing parameter $\xi$, the
structure of this expansion itself does not depend on it. It
vanishes as the mass gap goes formally to zero, while the PT part
survives. The INP part of the full gluon propagator in Eq. (5.14)
is nothing but the corresponding Laurent expansion in integer
powers of $q^2$ accompanied by the corresponding powers of the
mass gap squared and multiplied by the sum over the
$q^2$-independent factors, the so-called residues $\Phi_k(\lambda,
\alpha, \xi, g^2) = \sum_{m=0}^{\infty} \Phi_{k,m}(\lambda,
\alpha, \xi, g^2)$. The sum over $m$ indicates that an infinite
number of iterations (all iterations) of the corresponding
regularized skeleton loop integrals invokes each severe IR
singularity labelled by $k$. It is worth emphasizing that now this
Laurent expansion cannot be summed up into anything similar to the
initial Eq. (3.16), since its residues at poles gain additional
contributions due to the splitting/shifting procedure, i.e., they
become arbitrary. However, this arbitrariness is not a problem,
because severe IR singularities should be treated by the DRM
correctly implemented into the DT. For this the dependence of the
residues on their arguments is all that matters and not their
concrete values. The PT part of the full gluon propagator, which
has only the PT IR singularity, remains undetermined. This is the
price we have paid to fix the functional dependence of its INP
part. In Refs. \cite{10,16} we came to the same structure (5.14)
but in a rather different way. Concluding, it is worth emphasizing
that both terms in the general iteration solution (5.13) are valid
in the whole energy/momentum range, i.e., they are not
asymptotics. At the same time, we achieved the exact separation
between the two terms responsible for the NP (dominating in the IR
($q^2 \rightarrow 0$)) and the nontrivial PT (dominating in the UV
($q^2 \rightarrow \infty$)) dynamics in the true QCD vacuum. This
separation is unique as well, since for severe (i.e., NP) IR
singularities there exists a special regularization expansion
within the DT, complemented by the DRM, while for the PT IR
singularity it does not exist \cite{10,11}.

\section{Conclusions}

Our consideration at this stage is necessarily formal, since the
mass gap $\Delta^2$ remains neither IR (within the INP solution)
nor UV renormalized yet. At this stage it has been only
regularized, i.e., $\Delta^2 \equiv \Delta^2(\lambda, \alpha, \xi,
g^2)$. However, there is no doubt that it will survive both
multiplicative renormalization (MR) programs (which include the
corresponding removal of both $\lambda$ and $\alpha$ parameters).
For some preliminary aspects of the IRMR program see Refs.
\cite{10,16}. Anyway, how to conduct the UVMR program is not our
problem (as mentioned above, it is a standard one, and its
description can be found in Refs. \cite{3,4,6,7,8,9,14}). It is
worth noting that the mass gap which appears in the gluon SD
equation cannot be in principle the same one which appears in the
INP part of the general iteration solution, though we have
identified them, for simplicity.

It is important to emphasize that a mass gap has not been
introduced by hand. It is hidden in the skeleton loop integrals,
contributing to the gluon self-energy, and dynamically generated
mainly due to the NL interaction of massless gluon modes. No
truncations/approximations and no special gauge choice are made
for the above-mentioned regularized skeleton loop integrals. An
appropriate subtraction scheme has been applied to make the
existence of a mass gap perfectly clear. Within the general
iteration solution the mass gap shows up when the gluon momentum
goes to zero. The Lagrangian of QCD does not contain a mass gap,
while it explicitly appears in the corresponding gluon SD
equation. This once more underlines the importance of
investigation of the SD system of equations and identities
\cite{3,4,8,9} for understanding the true structure of the QCD
ground state. We have established the general structure of the
regularized full gluon propagator (see Eqs. (3.15) and (3.16)),
and the corresponding SD equation (3.12) in the presence of a mass
gap.

In order to realize a mass gap, we propose not to impose the
transversality condition on the gluon self-energy (see Eq. (3.5)),
while preserving the color gauge invariance condition (3.14) for
the full gluon propagator. This proposal is justified by the NL
and NP dynamics of QCD. Such a temporary violation of color gauge
invariance/symmetry (TVCGI/S) is completely NP effect, since in
the PT limit $\Delta^2=0$ this effect vanishes. Let us emphasize
that we would propose this even if there were no explicit
violation of the transversality of the gluon self-energy by the
constant skeleton tadpole term. In other words, whether this term
is explicitly present or not, but just color confinement (the
gluon is not a physical state) gives us a possibility not to
impose the transversality condition on the gluon self-energy. The
existence of this term is a hint that the above-mentioned
transversality might be temporary violated. Since gluon is not a
physical state because of color confinement as mentioned above,
the TVCGI/S in QCD has no direct physical consequences. None of
physical observables in QCD will be directly affected by this
proposal. For their calculations from first principles in
low-energy QCD we need the full gluon propagator, which
trasversality has been sacrificed in order to realize a mass gap
(despite their general role the ghosts cannot guarantee its
transversality in this case). However, we have already formulated
a general method how the transversality of the gluon propagator
relevant for NP QCD is to be restored at the final stage. In
accordance with our prescription \cite{17} it becomes
automatically transversal, free of the PT contributions
("contaminations"), and it regularly depends on the mass gap, so
that it vanishes when the mass gap goes to zero.

On the general ground (no truncations/approximations and no
special gauge have been made) we have established the existence at
least of two different types of solutions for the full gluon
propagator in the presence of a mass gap. The so-called general
iteration solution (5.14) is always severely singular in the IR
($q^2 \rightarrow 0$), i.e., the gluons always remain massless,
and this does not depend on the gauge choice (this behavior of the
full gluon propagator in different approximations and gauges has
been earlier obtained and investigated in many papers, see, for
example Ref. \cite{10} and references therein). The massive-type
solution (4.4) leads to an effective gluon mass, which explicitly
depends on the gauge-fixing parameter, and it cannot be directly
identified with the mass gap. Moreover, we were unable to make an
effective gluon mass a gauge-invariant as a result of the
renormalization, and therefore to assign to it a physical meaning.
This solution becomes smooth at $q^2 \rightarrow 0$ in the Landau
gauge $\xi=0$ only. Both types of solutions are independent from
each other and should be considered on equal footing, since the
gluon SD equation is highly NL system. For such kind of systems
the number of solutions is not fixed $a \ priori$. The UV behavior
($q^2 \rightarrow \infty$) of all solutions should be fixed by AF
\cite{3}. Due to unsolved yet confinement problem, the IR behavior
($q^2 \rightarrow 0$) is not fixed. Only solution of the color
confinement problem will decide which type of formal solutions
really takes place. At the present state of arts none of them can
be excluded.

In summary, the behavior of QCD at large distances is governed by
a mass gap, possibly realized in accordance with our proposal. The
dynamically generated mass gap is usually related to breakdown of
some symmetry (for example, the dynamically generated quark mass
is an evidence of chiral symmetry breakdown). Here a mass gap is
an evidence of the TVCGI/S. Thus there is no breakdown of $U(1)$
gauge symmetry in QED, since the photon is a physical state. At
the same time, a temporary breakdown of $SU(3)$ color gauge
symmetry in QCD is possible, since the gluon is not a physical
state (color confinement). In the presence of a mass gap the
coupling constant becomes play no role. This is also a direct
evidence of the "dimensional transmutation", $g^2 \rightarrow
\Delta^2(\lambda, \alpha, \xi, g^2)$ \cite{3,18,19}, which occurs
whenever a massless theory acquires masses dynamically. It is a
general feature of spontaneous symmetry breaking in field
theories. The mass gap has to play a crucial role in the
realization of the quantum-dynamical mechanism of color
confinement \cite{5}.

{\bf Acknowledgments.} Support in part by HAS-JINR and Hungarian
OTKA-T043455 grants (P. Levai) is to be acknowledged. The author
is grateful to A. Kacharava, N. Nikolaev, P. Forgacs, L. Palla, J.
Nyiri, and especially to C. Hanhart and A. Kvinikhidze for useful
discussions and remarks during his stay at IKP (Juelich).

\end{document}